\documentstyle[emulateapj,epsf,apjfonts]{article}

\slugcomment{To appear in Astrophysical Journal Letters}
\lefthead{Guinan et al.}
\righthead{Distance to the LMC}

\begin{document}

\title{\bf The Distance to the Large Magellanic Cloud from the Eclipsing Binary HV2274}

\author{E.F.~Guinan,~E.L.~Fitzpatrick,~L.E.~DeWarf,~F.P.~Maloney,~P.A.~Maurone}
\affil{Department of Astronomy \& Astrophysics, Villanova University, Villanova, PA 19085}

\author{I.~Ribas}
\affil{Departament d'Astronomia i Meteorologia, Universitat de Barcelona, \\ E-08028 Barcelona, Spain}

\author{J.D.~Pritchard}
\affil{Mount John University Observatory and Department of Physics \& Astronomy, \\ University of Canterbury, Private Bag 4800, Christchurch, New Zealand}

\author{D.H.~Bradstreet}
\affil{Department of Physical Science, Eastern College, St. Davids, PA 19087}

\author{A.~Gim\'enez}
\affil{Laboratorio de Astrofisica Espacial y Fisica Fundamental, Villafranca, Spain}

\begin{abstract}

The distance to the Large Magellanic Cloud (LMC) is crucial for the
calibration of the Cosmic Distance Scale.  We derive a distance to the
LMC based on an analysis of ground-based photometry and {\it HST}-based
spectroscopy and spectrophotometry of the LMC eclipsing binary system
HV2274.  Analysis of the optical light curve and {\it HST}/GHRS radial
velocity curve provides the masses and radii of the binary components.
Analysis of the {\it HST}/FOS UV/optical spectrophotometry provides the
temperatures of the component stars and the interstellar extinction of
the system.  When combined, these data yield a distance to the binary
system.  After correcting for the location of HV2274 with respect to
the center of the LMC, we find $d_{\rm LMC} = 45.7\pm1.6$ kpc or
$(V_0-M_v)_{\rm LMC}=18.30\pm0.07$ mag.  This result, which is immune
to the metallicity-induced zero point uncertainties that have plagued
other techniques, lends strong support to the ``short'' LMC distance
scale as derived from a number of independent methods.

\end{abstract}


\section{Introduction}

We present the first accurate distance determination to the Large
Magellanic Cloud (LMC) using an eclipsing binary system, resulting from
an international collaboration studying the physical properties of
these important objects.  This program includes both ground-based and
{\it HST} observations and is aimed primarily at determining the
stellar properties (masses, radii, luminosities, and ages) {\it and the
distances} of about a dozen selected $+14 < V < +16$ mag eclipsing
binaries in the Clouds (see Guinan et al. 1996).  It is well known that
studies of eclipsing binaries yield the most fundamental determinations
of the basic stellar properties.  We demonstrate here their great value
as ``standard candles'' for distance determination.

The distance to the LMC is a critical rung on the cosmic distance
ladder and numerous independent methods (utilizing, for example, RR
Lyrae stars, Cepheids, SN1987a, and ``red clump'' stars) have been
employed to determine it.   Unfortunately, and despite the use of the
eagerly-awaited {\it Hipparcos}\ parallaxes, there remains considerable
disagreement about this important measurement and, as summarized by Cole
(1998; see also Westerlund 1997), the current uncertainty is about
10-15\%.  In this paper, we show that the analysis of several eclipsing
LMC binaries could reduce the uncertainty in the LMC distance to level
of several percent.

These first results deal with the LMC eclipsing binary Harvard Variable
2274 (hereafter HV2274; $<V>_{max}\simeq14.2$; B1-2 IV-III + B1-2
IV-III; $P=5.73$ days).  This system is relatively bright, lightly
reddened, and has complete CCD light curves showing it to have deep
eclipses and uncomplicated out-of-eclipse light variations (Watson et
al. 1992).  Moreover, it is a detached system with nearly identical
bright stars located in an uncrowded field.  To provide an accurate
distance to the system and, hence, the LMC, absolute radii,
temperatures, and reddening corrections are needed (Guinan 1993;
Gim\'enez et al. 1994; Guinan et al. 1996). The radii are obtained from
the light and radial velocity curves and the temperatures and reddening
from UV/optical spectrophotometry.  The following sections present the
{\it HST} and ground-based observations used in this study (\S 2),
describe our analysis techniques (\S 3), and present our distance
determination for the LMC along with a discussion of the uncertainties
in this result (\S 4).

\section{Observations}

The ground-based $V$ band light curve data for HV2274 are taken from
Watson et al. (1992) and are shown in the lower panel of Figure 1
(``plus signs'').  The light curve is well-defined and shows two minima
of about equal depth ($\sim$0.7 mag) and nearly constant light outside
of the eclipses.  The more shallow minimum (secondary eclipse) occurs
at orbital phase 0.53, indicating an eccentric orbit ($e=0.136$).  We
also adopt the very recently available $V$ and $B-V$ measurements from
Udalski et al. (1998), i.e., $V = 14.16$ and $B-V = -0.13$.

The spectrophotometry and radial velocity curves needed to complete the
analysis of HV2274 were obtained by {\it HST}\ as part of our larger
project on eclipsing binaries in the Clouds.  Four FOS spectra,
covering four different wavelength regions, were acquired, calibrated,
and merged to produce a single spectrum spanning 1150~\AA~to 4820~\AA.
These observations were made at orbital phases outside of the eclipses,
when both stars were completely unobscured.

We also obtained 16 {\it HST}/GHRS medium resolution ($R=23000$)
spectra at a number of selected orbital phases to construct the radial
velocity curve.  The spectra covered 34~\AA~and were centered at two
different wavelengths: one near $\lambda$1305~\AA~and the other near
$\lambda$1335~\AA.  These regions contain strong UV photospheric lines
in early B stars (see Massa 1989).  Stellar \ion{Fe}{3}
$\lambda$1292.2, \ion{Si}{3} $\lambda\lambda$1300 triplets (1294.5,
1296.7 and 1298.9~\AA), and \ion{C}{2} $\lambda$1323.9 lines were fit
with double gaussians, which gave the radial velocities for each
component.  The individual line measurements for each component were
averaged to obtain its radial velocity at each phase.  In the upper
panel of Figure 1, these radial velocities are plotted against orbital
phase as determined by Watson et al. (1992; filled symbols).  The mean
uncertainties in the radial velocity measures are approximately
$\pm$15~km~s$^{-1}$.

\section{Analysis}
\subsection{Modeling the light and radial velocity curves}

The radial velocity and light curve data were analyzed using an
improved Wilson-Devinney light curve analysis code (Wilson \& Devinney
1971; Wilson 1990) that includes Kurucz atmosphere models for the
computation of the stellar radiative parameters (Milone et al. 1994).
The Wilson-Devinney code is a standard tool in the analysis of
eclipsing binaries.
  
The radial velocity and the light curve solutions for HV2274 are shown
in the top and bottom panels of Figure 1, respectively (solid curves).
For the light curve, the differences between observed and computed
values at $V$ (``(O-C)$_V$'') are also shown. Tables 1 lists the values
determined for the orbital period ({$P$}), eccentricity ({$e$}),
orbital inclination ({$i$}), longitude of periastron ($\omega$),
velocity semi-amplitudes ({$K$}), systemic velocity ($\gamma$),
temperature ratio ($T_B/T_A$), luminosity ratio ($L_B/L_A$), absolute
stellar radii ({$r$}), masses ({$M$}), and surface gravities ({$g$}).

The accurate measurement of the radii ({$r$}) of the components is
critically important for deriving the distance to HV2274.  The radii
are determined by combining the fractional radii ($r_f = r/a$) obtained
from the light curve analysis with the orbital semi-major axis ({\it
a}) derived chiefly from the spectroscopic orbit.  Thus, the absolute
radius of each star is $r = r_f \times a$ and the calculated
uncertainties in the radii are about 2.3\%.  The precision achievable
in the measurement of fundamental properties, such as radius, is
one of the keys to the usefulness of eclipsing binaries as standard
candles.

\subsection{Modeling the spectrophotometry}

The observed energy distribution of HV2274, $f_{\lambda\oplus}$, as
obtained by the FOS between 1150 \AA\/ and 4800 \AA\/ is shown in
Figure 2 (filled circles; the lower of the two full spectra).  The
observed fluxes depend both on the surface fluxes of the binary's
components and on the attenuating effects of distance and interstellar
extinction.  This relationship can be expressed as:
\begin{eqnarray} \label{basic1}
f_{\lambda\oplus} & =& \frac{1}{d^2}[r_A^2 F_{\lambda}^A + r_B^2 F_{\lambda}^B ] \times 10^{-0.4A(\lambda)}
\end{eqnarray}
\noindent or, substituting for the total extinction, $A(\lambda)$,
\begin{eqnarray} \label{basic2}
f_{\lambda\oplus} &=&\left(\frac{r_A}{d} \right)^2 [F_{\lambda}^A + (r_B/r_A)^2 F_{\lambda}^B] 
\\
& &  ~~\times 10^{-0.4 E(B-V) [k(\lambda-V) + R] \nonumber}
\end{eqnarray}
where $F_{\lambda}^i$ $\{i=A,B\}$ are the emergent fluxes at the
surfaces of the two components of the binary, the $r_i$ are the radii
of the components, and $d$ is the distance to the binary.  $A(\lambda)$
is the total extinction along the line of sight to the system at each
wavelength $\lambda$ and is expressed in Eq. 2 as a function of
$E(B-V)$, the normalized extinction curve $k(\lambda-V)\equiv
E(\lambda-V)/E(B-V)$, and the ratio of selective to total extinction in
the $V$ band $R \equiv A(V)/E(B-V)$.  Our analysis consists of modeling
the observed energy distribution of HV2274 via a non-linear least
squares procedure to derive values of the parameters $(r_A/d)^2$,
$F_{\lambda}^i$, $E(B-V)$, and $k(\lambda-V)$ in Eq. 2.  Ultimately,
the distance to the binary will be determined from $(r_A/d)^2$.

We represent the stellar surface fluxes $F_{\lambda}^i$ by ATLAS9
atmosphere models from R.L. Kurucz, which each depend on four
parameters: effective temperature ($T_{eff}$), surface gravity ($g$),
metallicity (normally $z$), and microturbulence velocity ($\mu$).  The
eight parameters needed to define the surface fluxes of the two binary
components are constrained by the results of the light and radial
velocity curve analyses and we adopt the effective temperature ratio
$T_B/T_A$ and $\log g$ values listed in Table 1.  In addition, we
assume that both components of the binary have the same metallicity and
the same microturbulence velocity.  The ratio of the stellar radii
$(r_B/r_A)^2$ in Eq. 2 is determined from the light and radial velocity
curve solutions (see Table 1), and we adopt the standard Galactic value
of $R = 3.1$ (see \S 4).

The normalized extinction curve $k(\lambda-V)$ is modeled using the
results of Fitzpatrick \& Massa (1990), who showed that the functional
form of the extinction wavelength dependence in the UV is tightly
constrained.  The parameters describing the shape of the curve are
determined from the fitting procedure.  The method of smoothly joining
UV and optical extinction curves is discussed by Fitzpatrick (1999).

This technique of solving simultaneously for both the stellar
parameters of a reddened star {\it and} the shape of the UV/optical
extinction curve is described in detail by Fitzpatrick \& Massa (1999;
FM99).  They demonstrate that the new ATLAS9 models can reproduce the
observed UV/optical continua of unreddened main sequence (and slightly
evolved) B stars {\em to a level consistent with the uncertainties in
currently available spectrophotometric data.}  They further show that
both the model atmosphere parameters and the wavelength dependence of
interstellar extinction can be extracted from analysis of UV/optical
spectra of reddened stars.  This is possible because the spectral
``signature'' of interstellar extinction is very different from the
``signatures'' of temperature, surface gravity, metallicity, or
microturbulence.

The dereddened energy distribution of HV2274 is shown as the upper
spectrum in Figure 2 (small filled circles with observational error
bars) superimposed with the best-fitting model (solid "histogram"
curve).  The inset shows a blowup of the Balmer jump region.  The
dereddened $V$ and $B$ data from Udalski et al. (1998), converted to
flux units, are shown by the large filled circles.  The excellence of
this fit (the reduced $\chi^2$ is close to 1) is illustrative of the
fits achieved by FM99 for B stars in general.  The parameters of the
fit are listed in Table 1.  For B-type stars, the ``metallicity''
measured from modeling the UV/optical flux is most heavily influenced
by absorption from iron-group elements in the UV region.  Therefore in
Table 1 we refer to the derived metallicity as $[{\rm Fe}/{\rm H}]$,
and the result is quite reasonable for a LMC star.  The uncertainties
indicated in Table 1 are 1-$\sigma$ {\em internal errors} and
incorporate the full covariance (or interdependence) of all the
parameters.  I.e, if any of the parameters is changed by $\pm$
1-$\sigma$, the total $\chi^2$ of the best fit --- after all the other
parameters are reoptimized -- increases by +1.  The small size of these
uncertainties is testament to the lack of covariance among the
parameters and the quality of the data.

The best-fit model found in this paper differs significantly from that
reported in an earlier version of this work (Guinan et al. 1998).  This
results entirely from the inclusion here of the $V$ and $B-V$
measurements from Udalski et al. (1998).  The earlier work incorporated
very uncertain optical photometry from Watson et al. (1992), but these
data were weighted so low that results were based essentially entirely
on the FOS data.  This had an important effect because, unless the
observations extend through the 4400 -- 5500 \AA\/ region, a possible
degeneracy exists between the best-fit values of $E(B-V)$ and the
normalized extinction curve $k(\lambda-V)$.  I.e.  very similar
relative extinction corrections, $E(B-V) \times k(\lambda-V)$, can be
obtained with a range of $E(B-V)$ values.  In Guinan et al. (1998) the
degeneracy region was found to be $E(B-V) \simeq 0.08 - 0.12$, with the
best fit given by $E(B-V) \simeq 0.08$ --- the solution reported in
that paper.  This result appeared reasonable since, combined with the
Watson et al. photometry ($B-V = -0.18$), it yielded an intrinsic color
of $(B-V)_0 \simeq -0.26$, compatible with the spectral type of
HV2274.  Using the more accurate (and thus more heavily weighted)
Udalski et al. photometry, the reddening degeneracy disappears and a
reasonable fit to the entire observed spectrum can only be achieved
with $E(B-V) = 0.12$ --- a value at the opposite end of the degeneracy
``valley'' from that reported in Guinan et al. (1998), as expected from
Murphy's Law.

\section{The Distance to the LMC}

The distance to HV2274 is computed from the values of $(r_A/d)^2$ and
$r_A$ derived above, and the uncertainty in this result is found by
propagating the uncertainties in each quantity.  The total uncertainty
in $(r_A/d)^2$ has internal and external components.  The internal
1-$\sigma$ fitting error is listed in Table 1.  The external effects
include uncertainty in the the extinction ratio $R$, which we take to
be 1-$\sigma(R) = \pm$0.3, and uncertainty in the zero point of the FOS
spectrophotometry, which we assume to be 1-$\sigma(FOS)=\pm$2-3\%
(Bohlin 1996; Bless \& Percival 1996).  Quadratically combining these
independent influences yields a total uncertainty of $\sigma[(r_A/d)^2]
\simeq 4.5$\%.   The uncertainty in $r_A$ is taken from the
Wilson-Devinney analysis and is listed in Table 1.  From these results
we derive  $d_{\rm HV2274} = 46.8\pm1.6~{\rm kpc}$, or, $
(V_0-M_v)_{\rm HV2274}=18.35\pm0.07$ mag.  This result is 0.14 mag
smaller than that reported by Guinan et al. (1998) due to the
resolution of the aforementioned reddening degeneracy.

To obtain the distance to the center of the LMC, the above result must
be corrected for the position of HV2274 in the LMC.  The line-of-sight
to the HV2274 binary system projects onto the end of the western region
of the central bar of the LMC. The {\it HST}/GHRS spectra show strong
ISM lines from LMC gas in \ion{O}{1} $\lambda$1302.2 \AA\/ and
\ion{Si}{2} $\lambda$1304.4 \AA. These lines show only one component
with a Heliocentric radial velocity of 285~km~s$^{-1}$, similar to the
HV2274 systemic velocity of 312~km~s$^{-1}$.  These data suggest that
HV2274 lies in or close to the disk of the LMC.  To transform the
distance of HV2274 to the LMC optical center we assumed the position
angle of the line of nodes is $168^{\circ}$ and that the disk
inclination is $38^{\circ}$ (Schmidt-Kaler \& Gochermann 1992).  Simple
trigonometry then implies that HV2274 is located on the far-side of the
LMC, about 1100~pc behind the center. The distance of the center of the
LMC is then $d_{\rm LMC} = 45.7\pm1.6$ kpc or $(V_0-M_v)_{\rm
LMC}=18.30\pm0.07$ mag.

This result lends strong support to the ``short'' LMC distance scale
(see Cole 1998).  Its importance, however, lies not only in the value
of the distance itself,  but in the precision that appears possible
from using eclipsing binaries.  This precision can be understood by
noting two facts:  1) the values of $\log g$ for the two binary members
are {\em very} well determined by the light and radial velocity curve
analysis; and 2) the best fitting model atmosphere reproduces the
HV2274 energy distribution over the entire observed range (1200 -- 5500
\AA) to within the small observational uncertainties (see Fig. 2).
These imply that the stellar temperature must --- like the gravity ---
be well-determined, since features such as the Balmer jump are
sensitive to both properties.  Further, a correct temperature could not
be found unless the distorting effects of interstellar extinction were
properly estimated and removed, since the possibility of a degeneracy
in the reddening solution has been eliminated by the inclusion of $V$
band data in the analysis (\S 3.2) .  Finally, we note that this
technique is immune to the metallicity-induced zero point uncertainties
that have plagued many other LMC distance estimates because the
determination of the stellar metallicity is an explicit part of the
analysis and because the derived distance is actually extremely
insensitive to the metallicity.

The result from a single binary system --- as precise as it appears to
be --- does not by itself resolve the LMC distance issue.
Unanticipated external systematic effects, such as a peculiar location
of the star within the LMC, could compromise the derived distance.
Such effects can only be addressed through the analysis of more
systems, preferably with a variety of locations within the LMC and
covering a range of stellar properties.  The ultimate precision of the
LMC distance will best be evaluated by the range of results derived
from such analyses.  The results presented here, however, demonstrate
the power of eclipsing binaries to address fundamental astrophysical
issues, such as the derivation of basic stellar properties, and
cosmologically important issues, such as the LMC distance.

\acknowledgements

This work was supported by NASA grants NAG5-7113 and HST GO-06683,
and NSF/RUI AST-9315365. 



\begin{deluxetable}{ll}
\tablenum{1}
\tablewidth{0pc}
\tablecaption{Orbital and Stellar Parameters}
\startdata
 & \nl
\tableline
\multicolumn{2}{c}{\it Orbital Solution and Stellar Properties\tablenotemark{a}}   \nl
\nl
$P$ = 5.726006(12) days                     & $e = 0.136(12)$                         \nl
$i$ = 89$^{\circ}\!\!.6(1.3)$               & $a = 38.58(93)$ $R_{\odot}$             \nl
$\omega$(1990.6) = 73$^{\circ}\!\!.3(1.5)$  & $\gamma$ = +312(4) km s$^{-1}$  \nl
K$_A$ = 166.2(5.9) km s$^{-1}$              & K$_B$ = 177.3(5.8) km s$^{-1}$          \nl
$T_{B}/T_{A}=1.005(5)$                      & $L_B/L_A (V) = 0.844(5)$                \nl
$r_A = 9.84(24) R_{\odot}$                  & $r_B = 9.03(20) R_{\odot}$              \nl
$M_A = 12.1(4) M_{\odot}$                   & $M_B = 11.4(4) M_{\odot}$               \nl
$\log g_A = 3.54(3)$                        & $\log g_B = 3.58(3)$                    \nl
\tableline
\multicolumn{2}{c}{\it Stellar Properties from UV/Optical Spectrophotometry\tablenotemark{b}} \nl
\nl
$T_A = 23000(180) K$                        & $T_B = 23110(180) K$                    \nl
$[{\rm Fe}/{\rm H}]_{AB} = -0.45$(6)        & $\mu_{AB} = 1.9(7)$ km s$^{-1}$         \nl
$E(B-V) = 0.120$(9) mag                     & $(r_A/d)^2 = 2.249(63)\times10^{-23}$   \nl
\tableline
\multicolumn{2}{c}{\it Distance to HV2274} \nl                                       \nl
$d = 46.8(1.6)$~kpc                         & $V_0\!-M_v\!=\!18.35(0.07)$mag          \nl
\tableline
\multicolumn{2}{c}{\it Distance to Center of LMC} \nl                                     \nl
$d = 45.7(1.6)$~kpc                         & $V_0\!-\!M_v\!=\!18.30(0.07)$~mag       \nl
\tablenotetext{a}{The uncertainties in the parameters resulting from
the light curve and radial velocity curve analyses were set at three
times the r.m.s. scatter of several solutions obtained from different
initial conditions.}
\tablenotetext{b}{The uncertainties in the parameters derived from
analysis of the UV/optical spectrophotometry are 1-$\sigma$ internal
errors derived from a non-linear least squares fitting procedure (see
\S3.2).}
\enddata
\end{deluxetable}


\begin{figure*}
\epsscale{0.80}
\plotone{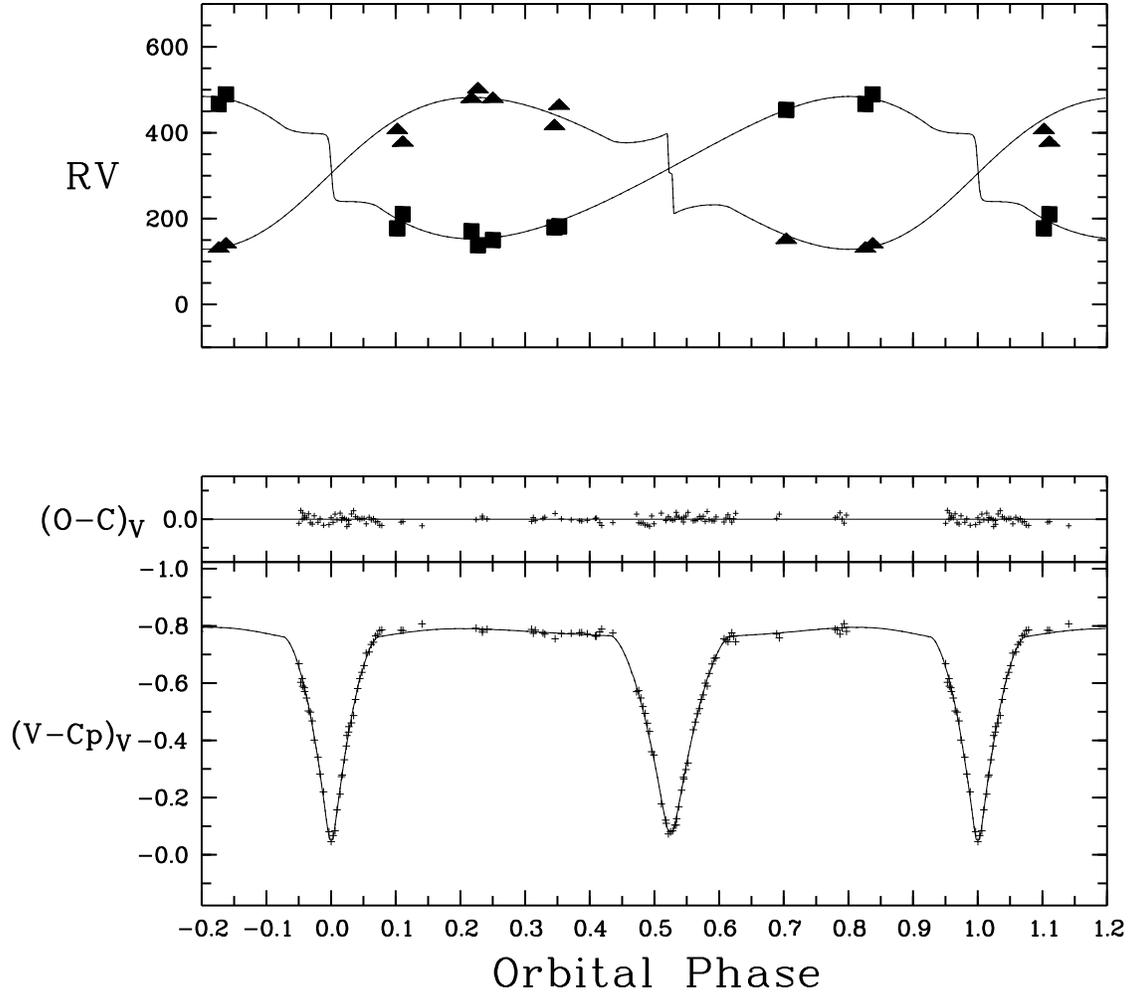}
\figcaption[guinan1.eps]{The upper panel shows the radial velocity
measurements of the two components of the HV2274 binary system as
derived from medium resolution {\it HST}/GHRS spectra (filled symbols)
and the radial velocity curve solution derived from the Wilson-Devinney
analysis (solid curves).  The discontinuous jumps seen in the model curves,
known as the ``Rossiter effect,'' occur when rotating stars are
partially eclipsed.  The lower panel shows the ground-based light curve
data from Watson et al. 1992,  based on differential photometry
(Variable $-$ Comparison) in the $V$ band (``V-Cp''; plus signs).  The
fit to the light curve is shown (solid curve) as well as the Observed
$-$ Computed  residuals (``O-C''). \label{fig:lcurve1}}
\end{figure*}

\begin{figure*}
\epsscale{0.90}
\plotone{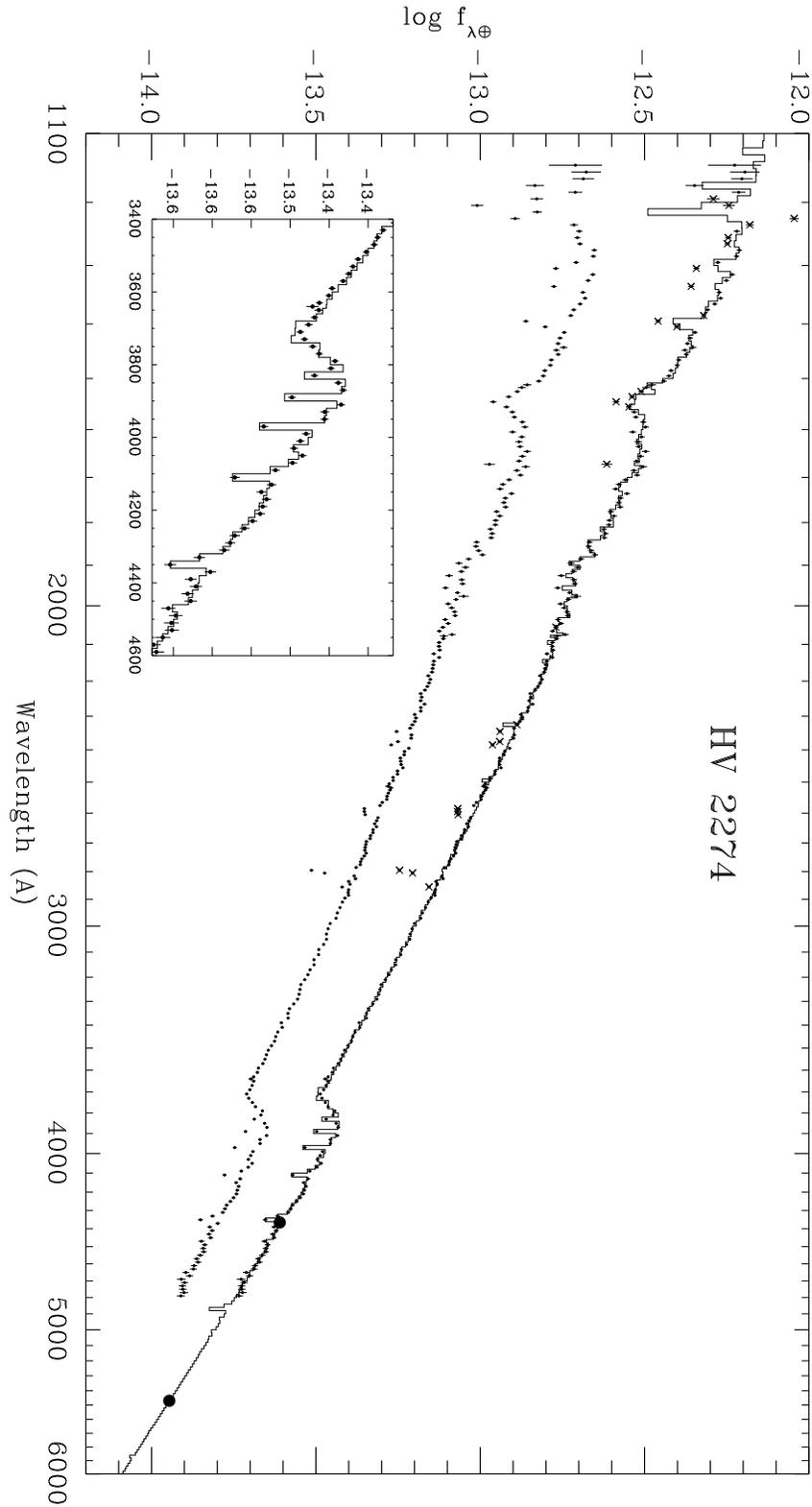}
\figcaption[guinan2.eps]{The UV/optical energy distribution of HV2274,
in units of ergs cm$^{-2}$s$^{-1}$\AA$^{-1}$.  The lower full spectrum
shows the observed HST/FOS energy distribution; the upper spectrum
shows the extinction-corrected energy distribution, superimposed with
the best-fitting ATLAS9 atmosphere model (plotted in histogram style).
Vertical lines through the data points indicate the 1-$\sigma$
observational errors.  Crosses indicate data points excluded from the
fit due to contamination by interstellar absorption lines.  The large
filled circles show dereddened B and V photometry from Udalski et al.
1998.  An expanded view of the fit near the Balmer jump is shown within
the inset.\label{fig:fosfit}}
\end{figure*}

\end{document}